\documentclass[
    a4paper            
  ]
  {jpconf}

\usepackage{epsfig}
\usepackage{tabularx}
\usepackage{graphicx}
\usepackage{amssymb,amsfonts,amsmath}

\begin{document}

\title{Direct neutralino searches in the NMSSM with gravitino LSP in the degenerate scenario}


\author{Grigoris Panotopoulos}

\address{Departament de Fisica Teorica, Universitat de Valencia, E-46100, Burjassot, Spain, and 
Instituto de Fisica Corpuscular (IFIC), Universitat de Valencia-CSIC,
Edificio de Institutos de Paterna, Apt. 22085, E-46071, Valencia, Spain}

\ead{grigoris.panotopoulos@uv.es}

\newcommand{\postscript}[2]{\setlength{\epsfxsize}{#2\hsize}
   \centerline{\epsfbox{#1}}}
\newcommand{\mweak}{M_{\text{weak}}}
\newcommand{\mplanck}{M_{\text{Pl}}}
\newcommand{\mstar}{M_{*}}
\newcommand{\OmegaM}{\Omega_{\text{M}}}
\newcommand{\OmegaDM}{\Omega_{\text{DM}}}
\newcommand{\ifb}{\text{fb}^{-1}}
\newcommand{\ev}{\text{eV}}
\newcommand{\kev}{\text{keV}}
\newcommand{\mev}{\text{MeV}}
\newcommand{\gev}{\text{GeV}}
\newcommand{\tev}{\text{TeV}}
\newcommand{\pb}{\text{pb}}
\newcommand{\mb}{\text{mb}}
\newcommand{\cm}{\text{cm}}
\newcommand{\km}{\text{km}}
\newcommand{\g}{\text{g}}
\newcommand{\s}{\text{s}}
\newcommand{\yr}{\text{yr}}
\newcommand{\sr}{\text{sr}}
\newcommand{\eg}{{\em e.g.}}
\newcommand{\ie}{{\em i.e.}}
\newcommand{\ibid}{{\em ibid.}}
\newcommand{\eqsref}[2]{Eqs.~(\ref{#1}) and (\ref{#2})}
\newcommand{\secref}[1]{Sec.~\ref{sec:#1}}
\newcommand{\secsref}[2]{Secs.~\ref{sec:#1} and \ref{sec:#2}}
\newcommand{\figref}[1]{Fig.~\ref{fig:#1}}
\newcommand{\figsref}[2]{Figs.~\ref{fig:#1} and \ref{fig:#2}}
\newcommand{\sla}[1]{\not{\! #1}}

\newcommand{\mB}{m_{B^1}}
\newcommand{\mq}{m_{q^1}}
\newcommand{\mf}{m_{f^1}}
\newcommand{\mKK}{m_{KK}}
\newcommand{\WIMP}{\text{WIMP}}
\newcommand{\SWIMP}{\text{SWIMP}}
\newcommand{\NLSP}{\text{NLSP}}
\newcommand{\LSP}{\text{LSP}}
\newcommand{\mWIMP}{m_{\WIMP}}
\newcommand{\mSWIMP}{m_{\SWIMP}}
\newcommand{\mNLSP}{m_{\NLSP}}
\newcommand{\mchi}{m_{\chi}}
\newcommand{\mgravitino}{m_{\gravitino}}
\newcommand{\mmed}{M_{\text{med}}}
\newcommand{\gravitino}{\tilde{G}}
\newcommand{\Bino}{\tilde{B}}
\newcommand{\photino}{\tilde{\gamma}}
\newcommand{\stau}{\tilde{\tau}}
\newcommand{\snu}{\tilde{\nu}}
\newcommand{\squark}{\tilde{q}}
\newcommand{\epsEM}{\varepsilon_{\text{EM}}}

\newcommand{\rem}[1]{{}}

\def\tauLbar{\overline {\tau_L}}
\def\Real{\Re e\,}
\def\Imag{\Im m\,}
\def\mg{m_{\tilde{G}}}
\def\msl{m_{\tilde{l}}}
\def\pstau{p_{\tilde{l}}}
\def\ptau{p_{{l}}}
\def\pZ{p_V}
\def\pg{p_{\tilde{G}}}

\begin{abstract}
In the present work a two-component dark matter model is studied adopting
the degenerate scenario in the R-parity conserving NMSSM. 
The gravitino LSP and the neutralino NLSP are extremely degenerate in mass, avoiding
the BBN bounds and obtaining a high reheating temperature for thermal leptogenesis.
In this model both gravitino (absolutely stable) and neutralino (quasi-stable)
contribute to dark matter, and direct detection searches for neutralino are discussed.
Points that survive all the constraints correspond to a singlino-like neutralino.
\end{abstract}


\section{Introduction}

There is accumulated evidence both from astrophysics and cosmology
that about 1/4 of the energy budget of the universe consists of so
called dark matter, namely a component which is non-relativistic and
neither feels the electromagnetic nor the strong interaction. For a
review on dark matter see e.g.~\cite{Munoz:2003gx}. Although the
list of possible dark matter candidates is long (for a nice list 
see~\cite{Taoso:2007qk}), it is fair to say
that the most popular dark matter candidate is the lightest
supersymmetric particle (LSP) in supersymmetric models with R-parity
conservation~\cite{Feng:2003zu}. For supersymmetry and supergravity see~\cite{susy}.
The simplest supersymmetric extension of
the standard model that solves the mu problem~\cite{Kim:1983dt} is the next-to-minimal 
supersymmetric standard model (NMSSM)~\cite{Nilles:1982dy}. If we do not consider
the axion~\cite{axion} and the axino~\cite{axino}, the superpartners that have the
right properties for playing the role of cold dark matter in the
universe are the gravitino and the lightest neutralino.
By far the most discussed case in the literature is the case of the
neutralino (see the classic review~\cite{Jungman:1995df}), probably
because of the prospects of detection. However, in the case is which
neutralino is assumed to be the only dark matter component, one has to face
the fine-tuning problem and the gravitino problem~\cite{Ellis:1982yb}. In most 
of the parameter space the neutralino relic density turns out to be either too small or 
too large~\cite{finetuning}. Furthermore, unstable gravitinos
will undergo late-time cascade decays to a neutralino LSP. These decays will destroy the 
light element abundances built up in BBN, unless $T_R < 10^5$~GeV~\cite{Kawasaki:2004yh},
which poses serious difficulties to the thermal leptogenesis scenario~\cite{Buchmuller:2002rq}.
If, on the other hand, gravitino is the LSP and therefore stable, playing the role of
cold dark matter in the universe, it is then the neutralino that will undergo late time decays
into gravitino and hadrons, and the gravitino problem is re-introduced~\cite{Feng:2004mt}.

It has been shown that in the degenerate scenario~\cite{Boubekeur:2010nt} the BBN and 
CMB constraints are avoided,
and high values of the reheating temperature are obtained compatible with thermal
leptogenesis. Here we focus on the scenario in which the masses of the gravitino LSP
and neutralino NLSP are extremely degenerate in mass. Under this assumption 
neutralino becomes quasi-stable participating to the cold dark matter of the universe 
together with gravitino, and it is still around and it can be seen in direct detection
searches experiments.

This article is organized as follows. In the next section we present
the theoretical framework. In section 3 we discuss all the relevant
constraints from colliders and from cosmology, and we show our
results. Finally, we conclude.

\section{Theoretical Framework}

\subsection{The NMSSM}

The particle physics model is defined by the superpotential
\begin{equation}\label{2:Wnmssm}
W= \epsilon_{ij} \left( Y_u \, H_2^j\, Q^i \, u + Y_d \, H_1^i\, Q^j
\, d + Y_e \, H_1^i\, L^j \, e \right) - \epsilon_{ij} \lambda \,S
\,H_1^i H_2^j +\frac{1}{3} \kappa S^3\,
\end{equation}
as well as the soft breaking masses and couplings
\begin{eqnarray}\label{2:Vsoft}
-\mathcal{L}_{\textrm{soft}}=&
 {m^2_{\tilde{Q}}} \, \tilde{Q}^* \, \tilde{Q}
+{m^2_{\tilde{U}}} \, \tilde{u}^* \, \tilde{u} +{m^2_{\tilde{D}}} \,
\tilde{d}^* \, \tilde{d} +{m^2_{\tilde{L}}} \, \tilde{L}^* \,
\tilde{L} +{m^2_{\tilde{E}}} \, \tilde{e}^* \, \tilde{e}
 \nonumber \\
& +m_{H_1}^2 \,H_1^*\,H_1 + m_{H_2}^2 \,H_2^* H_2 +
m_{S}^2 \,S^* S \nonumber \\
& +\epsilon_{ij}\, \left( A_u \, Y_u \, H_2^j \, \tilde{Q}^i \,
\tilde{u} + A_d \, Y_d \, H_1^i \, \tilde{Q}^j \, \tilde{d} + A_e \,
Y_e \, H_1^i \, \tilde{L}^j \, \tilde{e} + \textrm{H.c.}
\right) \nonumber \\
& + \left( -\epsilon_{ij} \lambda\, A_\lambda S H_1^i H_2^j +
\frac{1}{3} \kappa \,A_\kappa\,S^3 + \textrm{H.c.} \right)\nonumber \\
& - \frac{1}{2}\, \left(M_3\, \lambda_3\, \lambda_3+M_2\,
\lambda_2\, \lambda_2 +M_1\, \lambda_1 \, \lambda_1 + \textrm{H.c.}
\right) \,
\end{eqnarray}
When the singlet acquires a vaccum expectation value, S, we obtain an
effective $\mu$ parameter, $\mu_{eff}=\lambda S$. Imposing
universality at the GUT scale, a small controllable number
of free parameters remains, namely \\
\centerline{$tan \beta=v_u/v_d, m_0, A_0, m_{1/2}, \lambda, A_k$}
and the sign of the effective $\mu$ parameter can be chosen at will.

Because of the extra singlet superfield, in the NMSSM there is a
larger higgs sector and a larger neutralino sector.
The neutralino mass matrix is characterized by the appearence of
a fifth neutralino state, meaning that 
the composition of the lightest neutralino has an extra
singlino contribution
\begin{equation} \label{composition}
\tilde \chi^0_1 = N_{11} \tilde B^0 + N_{12} \tilde W_3^0 + N_{13}
\tilde H_1^0 + N_{14} \tilde H_2^0 + N_{15} \tilde S\,
\end{equation}
In the following, neutralinos with $N^2_{11}>0.9$, or
$N^2_{15}>0.9$, will be referred to as bino- or singlino-like,
respectively.

Furthermore, in the Higgs sector we have now two CP-odd neutral, and
three CP-even neutral Higgses. We make the assumption that there is
no CP-violation in the Higgs sector, and therefore the CP-even and
CP-odd states do not mix. We are not interested in the CP-odd
states, while the CP-even Higgs interaction and physical eigenstates
are related by the transformation
\begin{equation}\label{2:Smatrix}
h_a^0 = S_{ab} H^0_b\,
\end{equation}
where $S$ is the unitary matrix that diagonalises the CP-even
symmetric mass matrix, $a,b = 1,2,3$, and the physical eigenstates
are ordered as $m_{h_1^0} < m_{h_2^0} < m_{h_3^0}$.

\subsection{Production of gravitinos}

In the usual case (not in the degenerate scenario) gravitinos can be produced after inflation
in two ways. One way to produce gravitinos is with scatterings from the
thermal bath, and another is from the out-of-equillibrium decays of
the NLSP, which decouple from the thermal bath before primordial
Big-Bang Nucleosynthesis and decay after the BBN time. Thus, imposing
the WMAP bounds~\cite{Komatsu:2008hk} we can write for the gravitino abundance
\begin{equation}
0.1097 < \Omega_{3/2}h^2=\Omega_{3/2}^{TP}h^2+Y_{3/2}^{NLSP}h^2 < 0.1165
\end{equation}
where
\begin{equation}
\Omega_{3/2}^{NLSP} h^2 = \frac{\mgravitino}{m_{NLSP}} \:
\Omega_{NLSP} h^2
\end{equation}
with $m_{NLSP}$ the mass of the NLSP, and $\Omega_{NLSP} h^2$ the
abundance the NLSP would have, had it not decayed into the
gravitino. The thermal contribution is given by (approximately for a
light gravitino, $\mgravitino \ll m_{\tilde{g}}$)~\cite{Bolz:2000fu}
\begin{equation}
\Omega_{3/2}^{TP} \simeq 0.27 \: \left ( \frac{T_R}{10^{10}~GeV}
\right ) \: \left ( \frac{m_{\tilde{g}}}{TeV} \right )^{-2} \: \left
( \frac{\mgravitino}{100~GeV} \right )
\end{equation}
In the limit where $m_{NLSP} \rightarrow \mgravitino$ and $\tau_{NLSP} \gg 10^{17}~sec$
the scenario looks as if one would have a two-component dark matter with the NLSP
contribution $\Omega_{NLSP}h^2$, and a gravitino contribution from thermal production
only, $Y_{3/2}^{TP}$. Therefore, in the degenerate scenario with 
$m_{NLSP} \simeq \mgravitino$ the WMAP bound becomes
\begin{equation}
0.1097 < \Omega_{cdm} h^2=\Omega_{NLSP} h^2+\Omega_{3/2}^{TP}h^2 < 0.1165
\end{equation}
where from now on the NLSP is the lightest neutralino, $\chi=NLSP$.

\section{Constraints and results}

- Spectrum and collider constraints: We have used the computer software
NMSSMTools~\cite{Ellwanger:2004xm}, we have performed a random scan over the 
whole parameter space (with fixed $\mu > 0$ motivated by the muon
anomalous magnetic moment), and we have selected only those points
that satisfy all the theoretical requirements, as well as the LEP bounds on 
the Higgs mass, collider bounds on SUSY
particle masses, and experimental data from
B-physics~\cite{precision, Yao:2006px}. For all these good points
the lightest neutralino is either a bino or a singlino, and contrary
to the case where neutralino is the dark matter particle, here we do
not require that the neutralino relic density falls within
the allowed WMAP range.

- As we have already mentioned, the total dark matter abundance, and
not the neutralino one, should satisfy the cold dark matter
constraint~\cite{Komatsu:2008hk}
\begin{equation}
0.1097 < \Omega_{cdm} h^2=\Omega_{\chi} h^2+\Omega_{3/2}^{TP} h^2 < 0.1165
\end{equation}
that relates the reheating temperature after inflation to the
gravitino mass as follows
\begin{equation}
0.11 = A(\mgravitino, m_{\tilde{g}}) T_R+\Omega_{\chi} h^2
\end{equation}
For a given point in the cNMSSM parameter space, the complete
spectrum and couplings have been computed, and we are left with two
more free parameters, namely the gravitino mass and the reheating
temperature after inflation. The gravitino mass is equal essentially to the
neutralino mass, and the precise value can be determined if we specify the neutralino
lifetime. In the discussion to follow we have used a neutralino lifetime 
$\tau=10^{26} \: sec$, although the results
are not sensitive to it, and the figures we have produced for different values of the 
lifetime cannot be distinguished. Finally, the reheating temperature after inflation is 
obtained from the cold dark matter constraint. The thermal production contribution cannot 
be larger than the total dark matter abundance, and for this we can
already obtain an upper bound on the reheating temperature
\begin{equation}
T_R \leq 4.1 \times 10^9 \left ( \frac{\mgravitino}{100~\textrm{GeV}} \right ) \: \left
( \frac{\textrm{TeV}}{m_{\tilde{g}}} \right )^2 \: GeV
\end{equation}
Assuming a gluino mass $m_{\tilde{g}} \sim 1$~TeV, we can see that for a
heavy gravitino, $\mgravitino \sim 100$~GeV, it is possible to obtain a reheating
temperature large enough for thermal leptogenesis.

- For neutralino NLSP in the degenerate scenario, the only decay mode is 
$\chi \to \gamma \gravitino$,
for which the decay width can be computed once the supergravity Largrangian is 
known~\cite{Cremmer:1982en}, and it is given by~\cite{Feng:2004mt, Ellis:2003dn}
\begin{equation}
\Gamma(\chi \to \gamma \gravitino)
=
\frac{| N_{11} \cos \theta_W + N_{12} \sin \theta_W |^2}
{48\pi M_*^2} \
\frac{m_{\chi}^5}{m_{\gravitino}^2}
\left[1 - \frac{m_{\gravitino}^2}{m_{\chi}^2} \right]^3
\left[1 + 3 \frac{m_{\gravitino}^2}{m_{\chi}^2} \right]
\label{neutralinogamma}
\end{equation}
where $M_*$ is the Planck mass, $m_{\chi}$ is the neutralino mass,
and $\theta_W$ is the weak angle. In the limit where the mass difference
$\Delta m \equiv m_\chi-\mgravitino$ is much lower than the masses themselves,
$\Delta m \ll m_\chi, \mgravitino$, the neutralino lifetime becomes
\begin{equation}
\tau=\frac{1.78 \times 10^{13}~sec}{|N_{11} \cos \theta_W + N_{12} \sin \theta_W|^2} \: 
\left( \frac{GeV}{\Delta m} \right )^3
\end{equation}
From this formula one can see that for a mostly bino-neutralino a mass difference of 1 MeV 
is already enough to give a neutralino lifetime larger than the age of the universe.

- Neutralino-Nucleon spin-independent cross-section: LHC is now running and collecting data.
Although LHC is a powerful machine to look for physics beyond the standard model, it is 
known that other facilities are also needed to offer
complementary information towards the direction of searching for supersymmetry and 
identifying dark matter. The gravitino interactions are suppressed by the Planck mass,
and therefore direct production of gravitinos at colliders and/or direct detection prospects
seem to be hopeless. On the other hand, for a weakly interacting neutralino there are existing 
as well as future experiments that put experimental limits on the nucleon-neutralino 
cross-section. The spin-independent cross-section is given by
\begin{equation}
\sigma_{\chi-N}=\frac{4 m^2_r}{\pi} \, f_N^2\,
\end{equation}
where $m_r$ is the Nucleon-neutralino reduced mass,
$m_r=m_N m_{\chi}/(m_N +m_{\chi})$,
and
\begin{equation}
\frac{f_N}{m_N}\,=\,
\operatornamewithlimits{\sum}_{q=u,d,s} f_{Tq}^{(N)}
\frac{\alpha_{q}}{m_q} + \frac{2}{27}  f_{TG}^{(N)}
\operatornamewithlimits{\sum}_{q=c,b,t}
\frac{\alpha_{3q}}{m_q} \,
\end{equation}
In the above, $f_{TG}^{(N)}= 1- \operatornamewithlimits{\sum}_{q=u,d,s}
f_{Tq}^{(N)}$, we have taken the following
values for the hadronic matrix elements~\cite{Ellis:2000ds}:
\begin{align}\label{3:hadronvalues}
f_{Tu}^{(p)}= 0.020 \pm 0.004\,, \ \ \ &
f_{Td}^{(p)}= 0.026 \pm 0.005\,, &
f_{Ts}^{(p)}= 0.118 \pm 0.062\,,
\nonumber \\
f_{Tu}^{(n)}= 0.014 \pm 0.003\,, \ \ \ &
f_{Td}^{(n)}= 0.036 \pm 0.008\,, &
f_{Ts}^{(n)}= 0.118 \pm 0.062\,.
\end{align}
and $\alpha_q$ is the coupling in the effective Lagrangian
\begin{equation}
\mathcal{L}_{\mathrm{eff}} =\alpha_{i} \,\bar{\chi} \,\chi \, \bar q_i\, q_i
\end{equation}
where $i=1,2$ denotes up- and down-type quarks, and the Lagrangian is
summed over the three quark generations. The coupling $\alpha_q$ can be decomposed
into two parts, $\alpha_q=\alpha_q^h+\alpha_q^{\tilde{q}}$, where the first term
is the t-channel exchange of a neutral Higgs (Fig.~1), while the second term is the s-channel
exchange of a squark (Fig.~2). The expressions for $\alpha_q$ is terms of the masses and 
couplings of the model can be found in~\cite{Cerdeno:2004xw}.

Our main results are summarized in the figures below. In Fig.~3 and Fig.~4 we show the 
Nucleon-neutralino spin-independent cross section (in $\textrm{cm}^2$) versus neutralino mass and 
lightest Higgs boson (in GeV) respectively. The blue region corresponds to a bino neutralino, 
while the green
region corresponds to a singlino neutralino, and the curves are the current experimental limits
from CDMS~\cite{Ahmed:2009zw}. According to our results the bino scenario is already ruled 
out, while in the singlino case the 
upper region can be probed by future experiments. In Fig.~5 we show the reheating temperature 
after inflation as a function of the neutralino/gravitino mass. The blue region corresponds 
to a bino, the blue points correspond to singlino, and finally the red points correspond 
to singlino with relatively high values of the cross-section, 
namely $\sigma_{\chi-N} > 10^{-47}~cm^2$. The largest
values of $T_R$ correspond to a bino, which is ruled out, and for the singlino with relatively
high values of cross-section we obtain a reheating temperature $T_R \simeq 5 \times 10^9 \: GeV$
for a neutralino/gravitino mass $m_{\chi} \simeq \mgravitino \simeq 200 \: GeV$. In the last
figure we show the ($m_0$-$m_{1/2}$) plane ($m_0$ and $m_{1/2}$ in GeV) for singlino points 
with a cross-section larger than $10^{-47}~cm^2$, or lower than $10^{-47}~cm^2$. We see that
$m_0$ is not larger than 600 GeV, and therefore future direct detection experiments cannot
probe a region of the parameter space which can neither be probed by LHC.

\begin{figure}
\centerline{\epsfig{figure=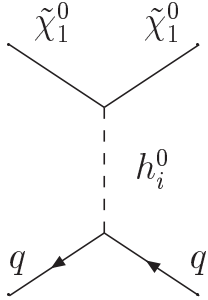,height=4cm,angle=0}}
\caption{Neutralino scattering off a nucleon by a neutral Higgs boson exchange.}
\end{figure}

\begin{figure}
\centerline{\epsfig{figure=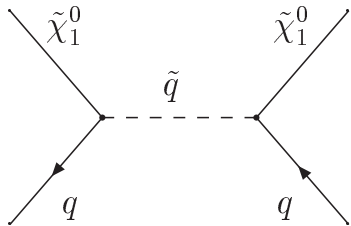,height=3cm,angle=0}}
\caption{Neutralino scattering off a nucleon by a squark exchange.}
\end{figure}

\begin{figure}
\centerline{\epsfig{figure=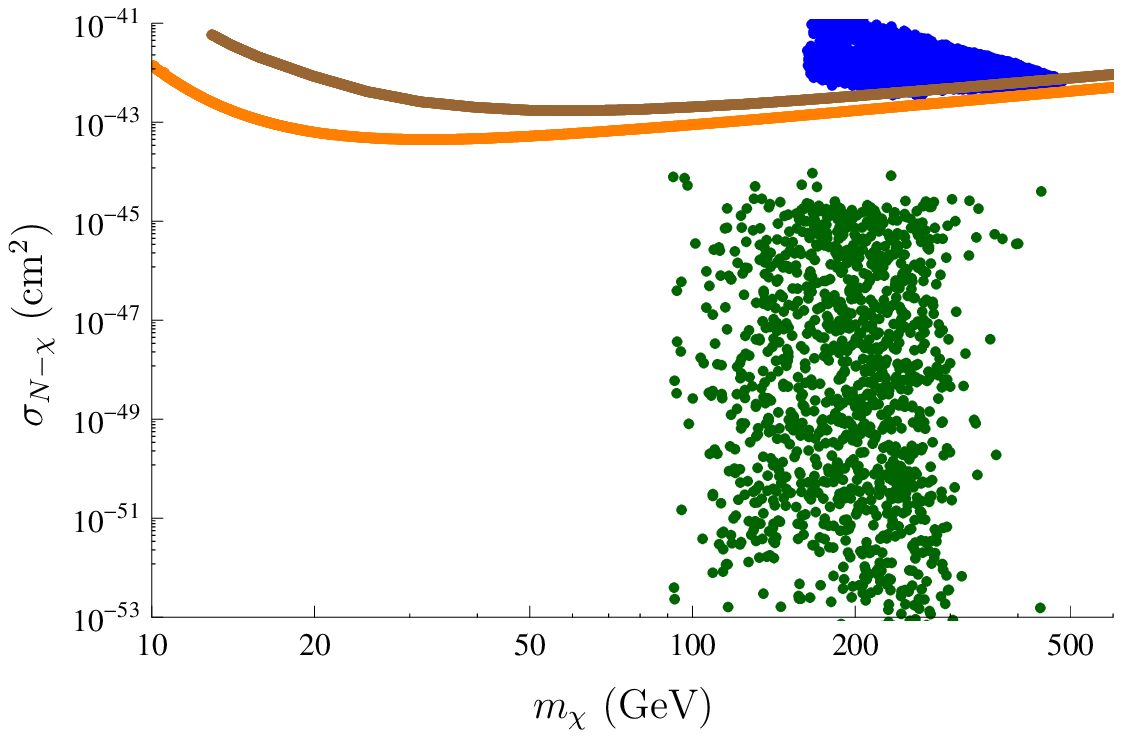,height=6cm,angle=0}}
\caption{Spin-independent neutralino-nucleon (proton) cross-section versus 
neutralino mass. Shown are the available experimental bounds from CDMS, and the 
predictions of the theoretical model. The blue region corresponds to a bino-like
neutralino, while the green points correspond to a singlino-like neutralino.}
\end{figure}

\begin{figure}
\centerline{\epsfig{figure=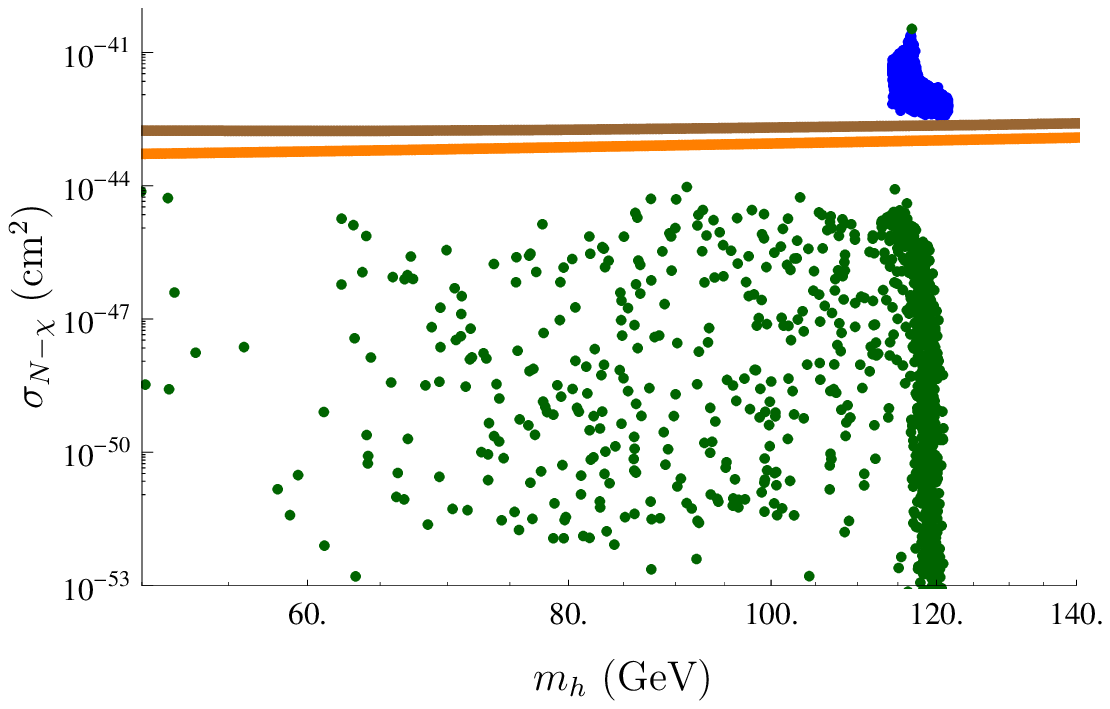,height=6cm,angle=0}}
\caption{Spin-independent neutralino-nucleon (proton) cross-section versus 
the lightest Higgs mass. Shown are the available experimental bounds CDMS, and 
the predictions of the theoretical model. The blue region corresponds to a bino-like
neutralino, while the green points correspond to a singlino-like neutralino.}
\end{figure}

\begin{figure}
\centerline{\epsfig{figure=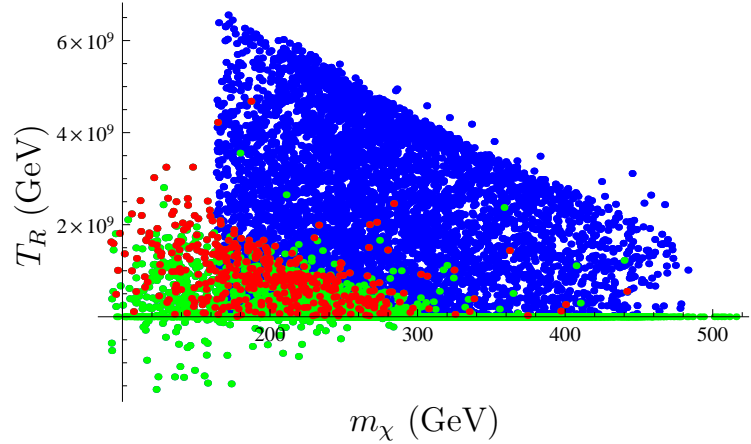,height=6cm,angle=0}}
\caption{Reheating temperature versus neutralino (or gravitino mass). Blue points
correspond to bino, green points correspond to singlino, and red points correspond
to singlino with a cross-section larger than $10^{-47}~cm^2$.}
\end{figure}

\begin{figure}
\centerline{\epsfig{figure=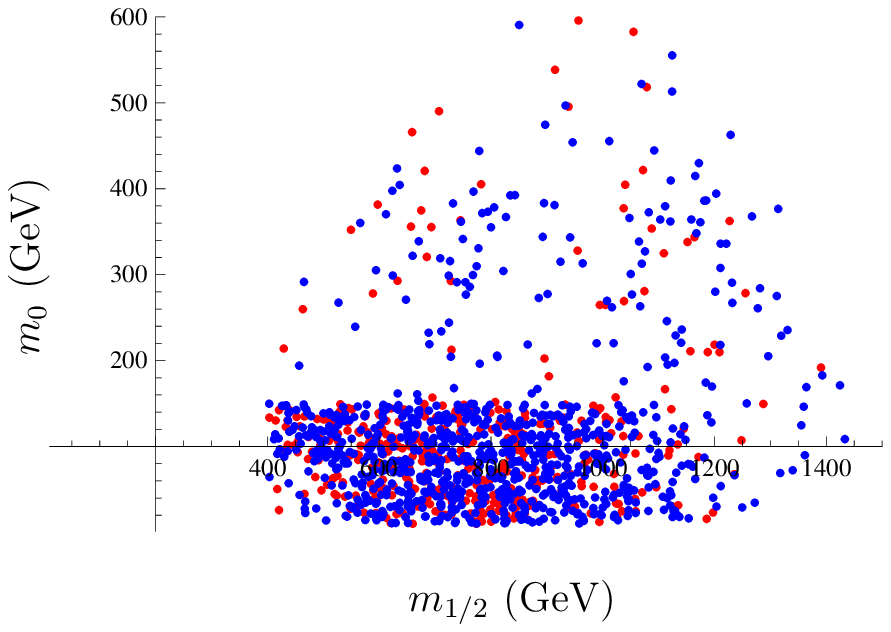,height=6cm,angle=0}}
\caption{The ($m_0$-$m_{1/2}$) plane for the singlino points. One color corresponds to
a cross-section larger than $10^{-47}~cm^2$, and the other color corresponds to
a cross-section lower than $10^{-47}~cm^2$.}
\end{figure}

\section{Conclusion}

In the framework of NMSSM, which solves the mu problem, we have assumed that the
gravitino LSP and the lightest neutralino NLSP are degenerate in mass.
Under this assumption the neutralino becomes extremely long-lived 
avoiding the BBN bounds. In this scenario we have a two component dark matter made
out of the absolutely stable gravitino and the quasi-stable neutralino. We have performed
a random scan over the whole parameter space keeping the points that satisfy the available
collider constraints plus the WMAP bound for dark matter. These points
correspond to either a bino or a singlino neutralino. We have computed the 
neutralino-nucleon spin-independent cross section as a function of the neutralino mass
and the lightest Higgs mass, and we find that the bino is ruled out. Then we explored
the ($m_0-m_{1/2}$) parameter space, and the reheating temperature dependence of the 
neutralino/gravitino mass for the singlino points that correspond to cross section values 
to be probed by future experiments.


\ack

We acknowledge financial support from FPA2008-02878, and Generalitat 
Valenciana under the grant PROMETEO/2008/004.

\end{document}